\documentclass{article}

\usepackage[preprint]{neurips_2026}

\usepackage[utf8]{inputenc} 
\usepackage[T1]{fontenc}    
\usepackage{hyperref}       
\usepackage{url}            
\usepackage{booktabs}       
\usepackage{amsfonts}       
\usepackage{nicefrac}       
\usepackage{microtype}      
\usepackage{xcolor}         
\usepackage{graphicx}
\usepackage{subcaption}

\workshoptitle{Representations for the Physical Sciences Workshop}
\title{Learning Radio Astronomical Representations with LeJEPA and Very Small Models}

\author{%
  Erica~Lastufka \\
  Department of Computer Science\\
    University of Geneva\\
  Geneva, Switzerland 1211 \\
  \texttt{erica.lastufka@unige.ch} \\
  \And
  Mariia Drozdova \\
  University of Geneva \\
  Geneva, Switzerland 1211 \\
  \texttt{mariia.drozdova@unige.ch} \\
  \And
  Vitaliy Kinakh \\
  University of Geneva \\
  Geneva, Switzerland 1211 \\
  \texttt{vitaliy.kinakh@unige.ch} \\
  \And
  Taras Holotyak \\
  University of Geneva \\
  Geneva, Switzerland 1211 \\
  \texttt{taras.holotyak@unige.ch} \\
    \And
  Miroslava Dessuages-Zavadsky \\
  University of Geneva \\
  Geneva, Switzerland 1211 \\
  \texttt{miroslava.dessuages@unige.ch} \\
    \And
  Daniel Schaerer \\
  University of Geneva \\
  Geneva, Switzerland 1211 \\
  \texttt{daniel.schaerer@unige.ch} \\
  \And
  Svyatoslav Voloshynovskiy \\
  University of Geneva \\
  Geneva, Switzerland 1211 \\
  \texttt{Svyatoslav.Voloshynovskyy@unige.ch} \\
}

\begin{document}

\maketitle

\begin{abstract}
Representations learned by vision foundation models pretrained on natural images have been shown to be useful for out-of-domain astronomical images. Performance on scientific downstream tasks increases with model size, which both carries higher inference costs and limits scalability, even when considering parameter-efficient adaptation. An alternative is to learn representations directly from astronomical observations rather than natural images, through self-supervised pretraining. 

We evaluate LeJEPA's ability to learn robust representations using very small vision models ($\sim$6M parameters) pretrained on Radio Galaxy Zoo images, comparing with established self-supervised frameworks. We test whether LeJEPA's latent-space regularization leads to better radio galaxy morphology classification. Across three evaluation datasets, LeJEPA achieves performance comparable to a substantially larger foundation model while producing more consistent representations across training and evaluation datasets. These results suggest that the choice of representation learning objective is critical for enabling small domain-specific models to achieve performance competitive with representations transferred from large foundation models in scientific imaging.
\end{abstract}

\section{Introduction}\label{sec:intro}

Modern foundation models continue to increase in size, with recent vision and vision-language models containing billions of parameters. Although larger models often achieve strong performance through broad pretraining, their scale makes both adaptation and inference increasingly expensive, even when using parameter-efficient methods such as LoRA \citet{hu_lora_2021}, which can still require training millions of parameters. Recent work has shown that such models can be successfully transferred to astronomical imaging tasks \citep{lastufka_examining_2025,drozdova_radio_2025}, suggesting that they encode broadly useful visual representations. However, this success comes at a significant computational cost: adapting and deploying large models requires substantial memory and compute resources, which can limit their scalability for scientific applications. This motivates investigating whether domain-specific self-supervised pretraining can achieve competitive representations with very small models.

Radio images present a particularly challenging domain for representation learning. Unlike optical galaxy images, they are reconstructed from interferometric measurements, contain substantially higher noise and dynamic range, and often exhibit smaller and less detailed sources. Self-supervised learning provides a way to learn representations directly from radio observations, reducing reliance on limited labeled datasets. However, it remains unclear whether representations learned from one radio dataset remain useful across a wide variety of same-domain datasets and tasks. Differences in observing configuration, including instrument, frequency, sensitivity, and target source populations can introduce dataset-specific variation that a representation may either retain or suppress. 

Motivated by the strong cross-domain representation learning with LeJEPA \citep{balestriero_lejepa_2025}, we investigate whether its predictive latent-space objective provides similar benefits for radio astronomical imaging. We compare LeJEPA with established contrastive and non-contrastive approaches, SimCLR \citep{chen_simple_2020} and BYOL \citep{grill_bootstrap_2020}, using small ($\sim$6M parameter) vision backbones trained on radio galaxy images. The learned representations are evaluated across radio morphology datasets using both linear probing and full fine-tuning, and their latent spaces are analyzed using embedding distribution metrics. 

\section{Data}\label{sec:tasks+data}

For model pre-training, a 20K sample of the Radio Galaxy Zoo (RGZ20k) dataset \citep{wong_radio_2025} was used. Radio images are galaxy-centered cutouts sourced from the NVSS and FIRST surveys of the VLA. Noise is suppressed in the images using sigma-clipping, where any pixels with a radio flux density of less than $3\sigma$ of the image mean are set to zero. It is noted in \citet{slijepcevic_can_2022} that many sources in this 20K sample are unresolved.

Evaluation via classification was performed using three currently available public datasets: MiraBest \citep{porter_mirabest_2023}, RadioGalaxyDataset \citep{griese_floriangrieseradiogalaxydataset_2022} which we refer to as FIRST, and the RGZ dataset from \citep{wu_radio_2019}. As MiraBest, FIRST, and RGZ20k all have been pre-processed in a similar manner with sigma-clipping, we decided to apply a linear min-max scaling to the RGZ images rather than keeping either scaling method (linear z-scale or log min-max) used by \citet{wu_radio_2019}.
FIRST also contains samples of MiraBest, as morphology labels are generally scarce so datasets build off of existing ones.

\begin{table*}[h]
\caption{\label{tab:datasets}Datasets used in this study.}
\centering
\resizebox{\textwidth}{!}{
\begin{tabular}{p{1.3cm}p{2cm}p{2cm}p{7cm}}
\toprule
Dataset & \raggedright{Samples} & Image size & Remarks  \\
\midrule
RGZ20k & 20K & 132$\times$132px & unlabeled, unbalanced \\
MiraBest & 934 & 150$\times$150px & FR-I and FR-II, balanced  \\
FIRST & 2158 & 300$\times$300px & FR-I, FR-II, compact, and bent-tailed, unbalanced \\
RGZ & 6.8K & 132$\times$132px & 6 classes, unbalanced \\

\hline
\end{tabular}}
\end{table*}

\section{Methods and Models}

We investigate whether LeJEPA's representation learning objective enables small self-supervised models trained on domain-specific unlabeled images to learn robust representations for radio astronomy. Unlike standard transfer learning, where a fixed encoder pretrained on a broad natural-image distribution is adapted to a scientific task, the encoder is learned directly from unlabeled radio data. The resulting representations are evaluated against a larger foundation model to assess whether domain-specific pretraining can provide competitive downstream performance at a much smaller scale. 
This comparison focuses on objective choice at matched backbone capacity, while view construction and optimization remain method-specific.

We evaluate several backbone architectures and self-supervised learning approaches. Our primary architectural comparison is between convolutional and transformer-based models using EfficientNet-B0 \citep{tan_efficientnet_2019} and ViT-Tiny \citep{dosovitskiy_image_2021}. These models have comparable parameter counts, with EfficientNet-B0 containing approximately 5.2M parameters and ViT-Tiny approximately 5.7M parameters, enabling a comparison of architectural inductive biases while controlling for model scale.

To evaluate whether LeJEPA's predictive latent-space objective provides advantages for radio representation learning, we compare it with two established self-supervised approaches using ViT-Tiny backbones: SimCLR and BYOL. SimCLR learns representations through contrastive instance discrimination, encouraging agreement between augmented views while separating different samples. BYOL uses a non-contrastive bootstrap approach, where an online network predicts the representation of a slowly updated target network. LeJEPA follows a joint-embedding predictive architecture, learning representations by predicting latent features while regularizing the latent distribution using SIGReg. This provides a direct comparison between contrastive, bootstrap, and predictive SSL objectives while controlling for model capacity.

Models are pretrained on RGZ20k for 200 epochs with a learning rate of $3\times10^{-5}$. A validation split containing 20\% of the training data is used to monitor pretraining convergence. LeJEPA is additionally evaluated using EfficientNet-B0 to determine whether any observed benefits extend beyond a single architecture. Full pre-training details are given in Appendix \ref{sec:app_pretrain}.

As external baselines, we include \citet{zhu_efficient_2026}'s Efficient Universal Perception Encoder (EUPE) ViT-Tiny and ViT-Small variants. These models provide a comparison against representations inherited from large-scale pretraining while remaining substantially smaller than their teacher models.

\section{Embedding-space distribution shifts}

\begin{figure}
    \centering
    \includegraphics[width=0.45\linewidth]{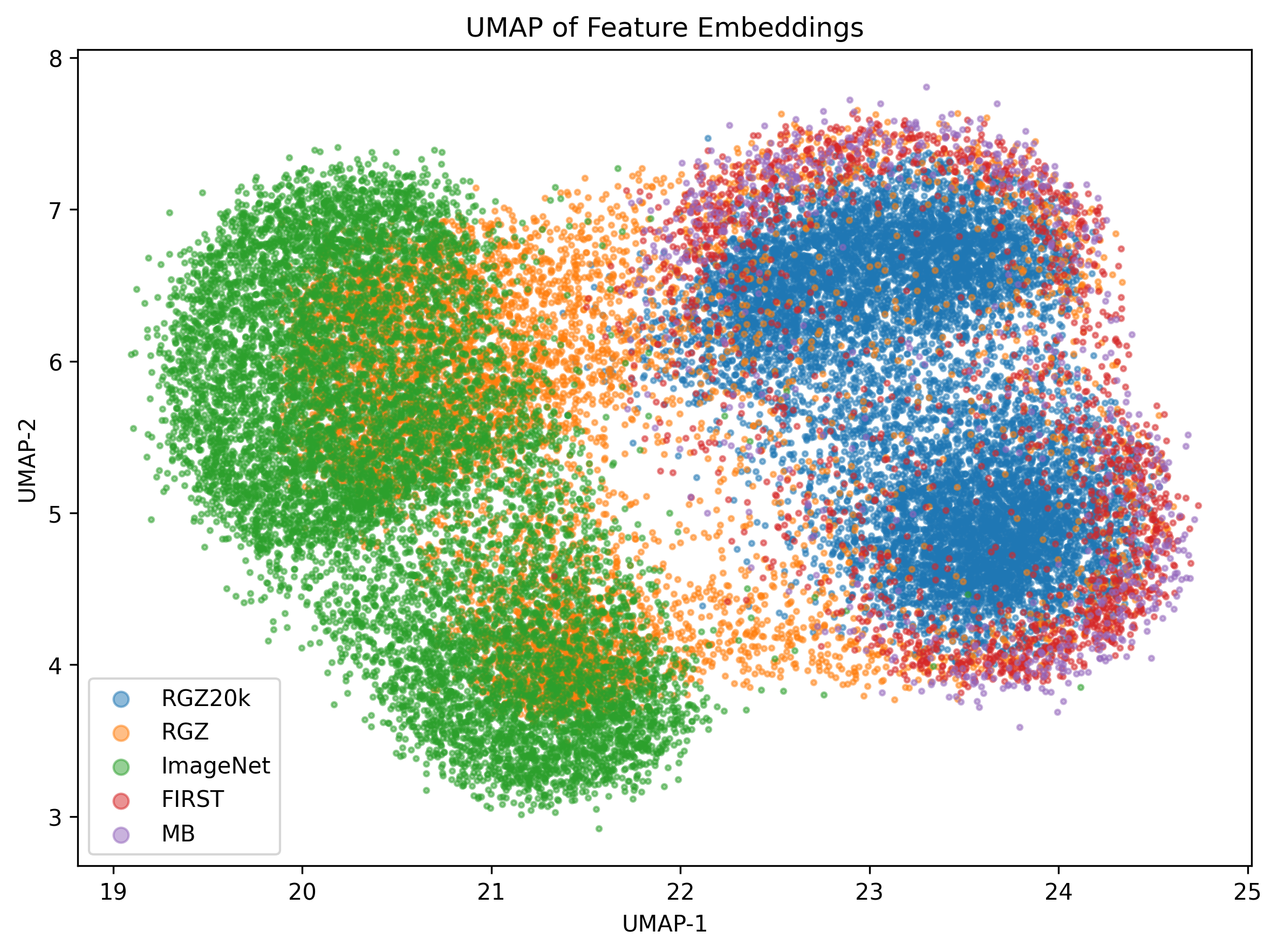}
    \includegraphics[width=0.45\linewidth]{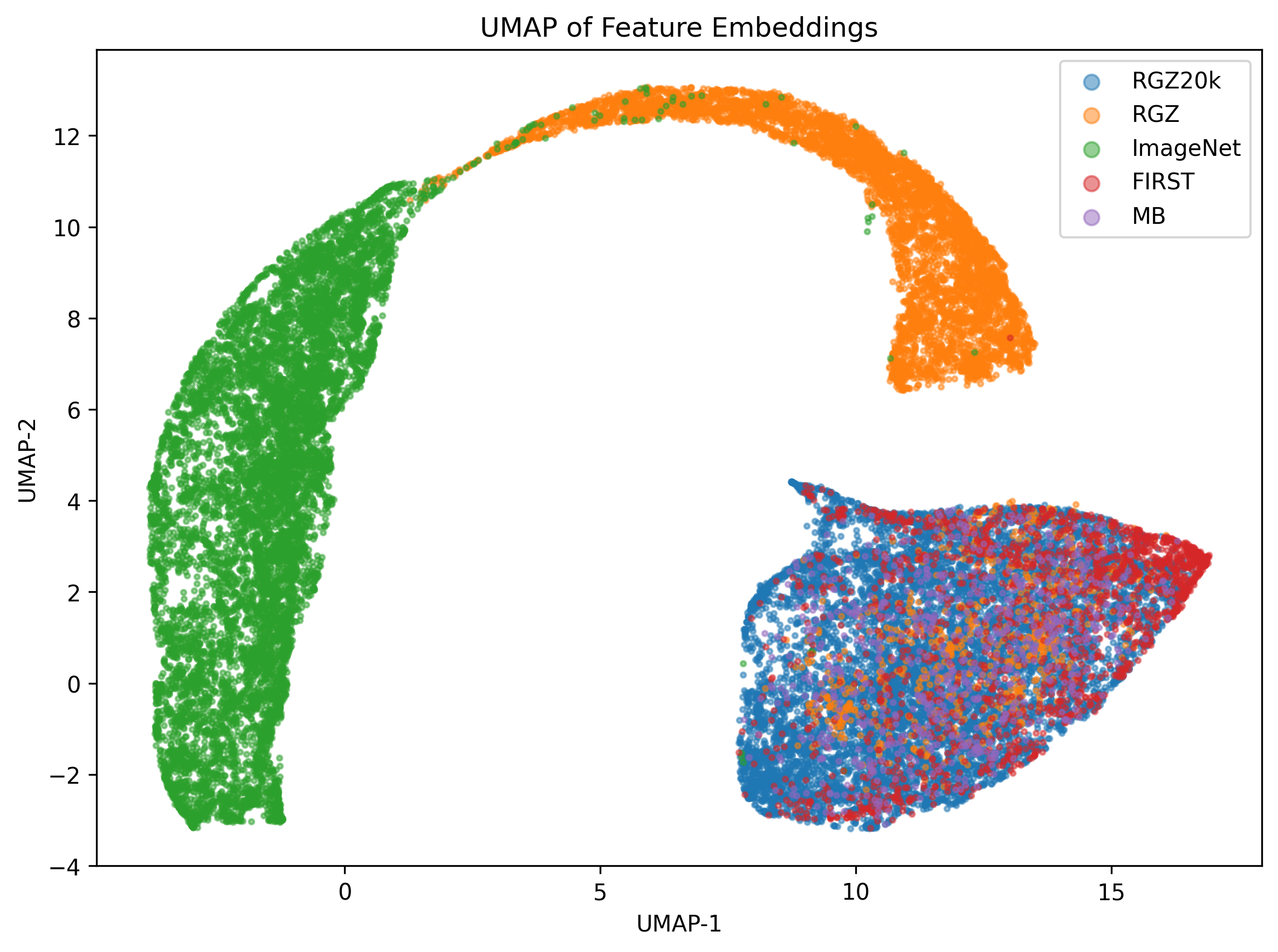}
    \caption{Two-dimensional UMAP \citep{mcinnes_umap_2020} projections of feature embeddings for our datasets, using EUPE-ViT-T (left) and LeJEPA ViT-Tiny (right). Other UMAPs are in Appendix \ref{sec:app_umaps}}
    \label{fig:umap}
\end{figure}

\begin{table}[b]
\centering
\caption{Distribution shift (MMD, with FD shown in parentheses) between datasets measured using different models as feature extractors. Lower values indicate more similar distributions.}
\label{tab:distribution_shift}
\resizebox{\linewidth}{!}{
\begin{tabular}{lrrrrrrr}
\toprule
 & RGZ20k--RGZ & RGZ20k--FIRST & RGZ20k--MB & IN--RGZ20k & IN--RGZ & IN--FIRST & IN--MB \\
\midrule
EUPE-ViT-T (6M) &
0.215 & -0.002 & 0.031 & 0.062 & -0.013 & 0.053 & 0.068 \\
&
(0) & (0) & (0) & (0) & (0) & (0) & (0) \\

EUPE-ViT-S (22M)&
0.374 & 0.127 & 0.148 & 0.450 & 0.404 & 0.421 & 0.400 \\
&
(12.542) & (2.336) & (3.740) & (127.337) & (128.919) & (126.012) & (122.423) \\


SimCLR ViT-T (6M)&
0.794 & 0.021 & 0.128 & 0.800 & 0.334 & 0.773 & 0.764 \\
&
(74.318) & (0.253) & (1.289) & (85.856) & (35.090) & (87.989) & (79.126) \\

BYOL ViT-T (6M)&
0.875 & 0.027 & 0.483 & 1.592 & 0.674 & 1.614 & 1.580 \\
&
(18.784) & (0.000) & (0.003) & (43.548) & (8.475) & (43.580) & (42.953) \\

LeJEPA ViT-T (6M)&
0.329 & 0.092 & 0.029 & 0.897 & 0.644 & 0.714 & 0.722 \\
&
(0.046) & (0.005) & (0.002) & (3.453) & (3.260) & (3.632) & (3.538) \\

LeJEPA EfficientNet (5M)&
0.324 & 0.039 & 0.208 & 0.439 & 0.097 & 0.373 & 0.301 \\
&
(2679.55) & (153.80) & (1503.15) & (3967.46) & (1962.54) & (3869.06) & (4533.60) \\
\bottomrule
\end{tabular}
}
\end{table}

LeJEPA pre-training immediately improves domain-specific representation quality with respect to an EUPE model of the same architecture (ViT-Tiny), as illustrated in Figure \ref{fig:umap}. While EUPE-ViT-T shows partial representation collapse, LeJEPA produces a more structured latent space: ImageNet (shown only for comparison) is clearly distinct from the various radio datasets, which themselves share many common features and thus occupy a similar region of the embedding space. This is also reflected in Table \ref{tab:distribution_shift}, where we compare the distributions produced by each encoder using Maximum Mean Discrepancy (MMD, \citet{gretton_kernel_2012}) and Fréchet distance (FD, \citet{heusel_gans_2018}). Both metrics are computed within the latent space of a given model and therefore measure differences between representations rather than the image distributions directly. 


A representation useful for transfer between radio datasets should capture features related to the underlying astronomical sources while being insensitive to differences that are specific to a particular instrument, observation, or preprocessing pipeline. Since the radio datasets used here contain similar types of astronomical objects and are derived from VLA images, we expect pre-training on RGZ20k to reduce the distribution shift between this and other radio datasets.  
Across all models, the measured shift between radio datasets is substantially smaller than the shift between radio observations and ImageNet. The exception is EUPE-ViT-T, whose FD values of 0 are consistent with representation collapse visible in Figure \ref{fig:umap}. LeJEPA further [mostly] reduces the distances between radio datasets compared with SimCLR and BYOL, while preserving the large separation from ImageNet. This suggests that LeJEPA learns representations that suppress some dataset-specific variation between radio observations without removing features that distinguish radio images from natural images.

 \section{Classification Performance}

The strength of each model's pre-trained representations was evaluated on the three classification datasets (RGZ, FIRST and MiraBest) via both full fine-tuning and a frozen backbone with a linear classifier. We selected appropriate learning rates between $1e-5$ and $1e-3$ for each model and used early stopping to prevent overfitting. Macro F1 scores for both full fine-tuning and linear probing are reported in Table \ref{tab:lejepa}, along with errors denoting variation across runs with 3 different random seeds. For EUPE-ViT-T, the frozen backbone did not produce representations separable by a simple linear classifier so no F1 score is reported. 


\begin{table}[h]
\resizebox{\textwidth}{!}{
    \centering
\begin{tabular}{lllllll}
\toprule
 & RGZ F1 & RGZ probe F1 & FIRST F1 & FIRST probe F1 & MB F1 & MB probe F1 \\
\midrule
LeJEPA ViT-T (6M)& \textbf{0.72 $\pm$ 7e-3} & 0.63 $\pm$ 1e-2 & \textbf{0.77 $\pm$ 1e-3} & 0.58 $\pm$ 5e-3 & \textbf{0.96} $\pm$ 1e-2 & 0.84 $\pm$ 5e-3 \\
LeJEPA EfficientNet (5M)& 0.71 $\pm$ 2e-3 & \textbf{0.68 $\pm$ 3e-3} & 0.76 $\pm$ 3e-3 & \textbf{0.73 $\pm$ 1e-2} & 0.94 $\pm$ 2e-2 & \textbf{0.93 $\pm$ 7e-3} \\
EUPE-ViT-S (22M)& 0.71 $\pm$ 4e-4 & 0.55 $\pm$ 3e-3 & 0.74 $\pm$ 3e-2 & 0.59 $\pm$ 9e-3 & 0.93 $\pm$ 1e-4 & 0.92 $\pm$ 1e-2 \\
SimCLR ViT-T (6M)& 0.60 $\pm$ 4e-3 & 0.25 $\pm$ 4e-3 & 0.66 $\pm$ 1e-2 & 0.64 $\pm$ 3e-2 & 0.83 $\pm$ 9e-3 & 0.76 $\pm$ 2e-2 \\
BYOL ViT-T (6M)& 0.65 $\pm$ 2e-2 & 0.67 $\pm$ 9e-3 & 0.54 $\pm$ 3e-2 & 0.38 $\pm$ 6e-2 & 0.76 $\pm$ 2e-3 & 0.76 $\pm$ 7e-2 \\
EUPE-ViT-T (6M)& 0.58 $\pm$ 3e-2 & - & 0.35 $\pm$ 7e-3 & - & 0.63 $\pm$ 1e-2 & - \\
\bottomrule
\end{tabular}
}
\caption{F1 classification score for all evaluation datasets, with both fine-tuning and linear probing.}
\label{tab:lejepa}
\end{table}

Results demonstrate that the self-supervised learning objective strongly influences the quality of learned representations for small models. LeJEPA pretraining consistently produces the strongest ViT-Tiny representations, as evidenced by its classification performance out-scoring SimCLR and BYOL across all three radio datasets when fine-tuned. After fine-tuning, LeJEPA ViT-Tiny achieves performance comparable to or even surpassing the larger 22M-parameter EUPE-ViT-S model, despite the substantially smaller model size. EfficientNet-B0 achieves the strongest probe performance on several datasets, indicating that its representations are well suited to the classification tasks, although this advantage is not as strong as ViT's after fine-tuning. Overall, the results show that when model capacity is held approximately constant, the choice of self-supervised objective can substantially affect the quality of the resulting representation. At approximately 6M parameters, LeJEPA produces more effective representations for radio astronomical images than the SimCLR and BYOL objectives. 

\section{Conclusion}

In this work, we investigated whether small self-supervised models trained on domain-specific unlabeled data can learn useful representations for radio astronomical imaging. Rather than relying on increasingly large foundation models, we studied the effect of the representation learning objective in a setting motivated by the computational constraints of scientific inference. By comparing LeJEPA with SimCLR and BYOL using similarly sized ViT-Tiny models, we found that the choice of learning objective has a substantial impact on downstream task performance.

LeJEPA consistently produced the strongest ViT-Tiny representations across three morphology classification datasets, outperforming contrastive and non-contrastive baselines and achieving comparable performance to a larger distilled foundation model, EUPE-ViT-Small. Analysis of the embedding distributions showed that different SSL objectives produce distinct latent geometries: while radio datasets derived from related VLA surveys exhibit relatively small distribution differences, these are amplified or suppressed depending on the learned representation. LeJEPA representations showed reduced sensitivity to variations between radio datasets, suggesting that predictive latent-space objectives may encourage  representations that are less dataset-specific.


These results show that the choice of self-supervised objective has a substantial effect on the representations learned by small scientific vision models. Across all experiments, LeJEPA provides the strongest overall performance. This makes predictive latent-space learning a strong candidate for building resource-efficient models for scientific imaging, offering a practical alternative to relying on increasingly large foundation models.

\newpage
\bibliographystyle{plainnat}
\bibliography{radio_lejepa}

\appendix

\newpage
\section{Appendix}

\subsection{Code}
Code for pre-training and evaluation is available at \url{https://anonymous.4open.science/r/radio_lejepa-E92D/}.

\subsection{Pre-training}\label{sec:app_pretrain}

\subsubsection{LeJEPA Pre-training}

All LeJEPA models were pre-trained on the RGZ20k dataset for 200 epochs using AdamW \citep{loshchilov_decoupled_2018} with a learning rate of $3\times10^{-5}$ and weight decay of $5\times10^{-2}$. The learning rate was linearly warmed up from $1\%$ of its initial value for the first epoch, followed by cosine annealing. Training used a batch size of 16, an input resolution of $224\times224$ pixels, and a fixed random seed.

For each image, two global views and eight local views were generated, following recommended best practices from the LeJEPA paper \citep{balestriero_lejepa_2025}. Global crops used a scale range of $[0.5,1.0]$, while local crops used $[0.4,0.7]$ and were resized to $224\times224$ after cropping. Larger local crops were chosen due to the sparse nature of the images. Standard image augmentations -- color jitter, Gaussian blur, horizontal flips, and ImageNet normalization -- were applied,  despite these not being necessarily the optimal augmentations for this particular dataset.

The LeJEPA objective combines an invariance loss with SIGReg:
\begin{equation}
\mathcal{L}_{\mathrm{LeJEPA}}
=
(1-\lambda)\mathcal{L}_{\mathrm{inv}}
+
\lambda\mathcal{L}_{\mathrm{SIGReg}},
\end{equation}
with $\lambda=0.05$. The projection head consists of an MLP with hidden dimensions $2048$ and output dimension $128$, with batch normalization. SIGReg used 17 integration knots and 256 randomly sampled projection directions. Training was performed using mixed-precision computation with bfloat16.

Pre-training completed faster with a ViT-Tiny backbone compared to an EfficientNet-B0 backbone. The lower inference latency of ViT-Tiny (8.55 ms vs. 13.33 ms for EfficientNet-B0) is consistent with its faster training, since each forward pass required less computation.

\subsubsection{BYOL Pre-training}

BYOL pre-training was performed on RGZ20k using a ViT-Tiny backbone with the same input resolution, augmentations, normalization, training duration, and batch size as described above, except that two augmented views were generated per image. The online and target networks used identical ViT-Tiny backbones with a 192-dimensional output. Each network was followed by a two-layer projector with dimensions $192\rightarrow512\rightarrow128$, while the online network additionally used a predictor with dimensions $128\rightarrow512\rightarrow128$. The online network and predictor were optimized jointly using AdamW with a learning rate of $3\times10^{-4}$ and weight decay of $10^{-6}$. The target network was updated using an exponential moving average with momentum $m=0.999$.

\subsubsection{SimCLR Pre-training}

SimCLR pre-training used the same augmentations and hyperparameters as BYOL pre-training. SimCLR used an NT-Xent (InfoNCE) loss with temperature $\tau=0.5$ and cosine annealing learning-rate scheduling. Mixed-precision training was used where enabled. 

\newpage

\subsection{UMAPs}\label{sec:app_umaps}

\begin{figure}[h]
    \centering
    \includegraphics[width=0.48\linewidth]
    {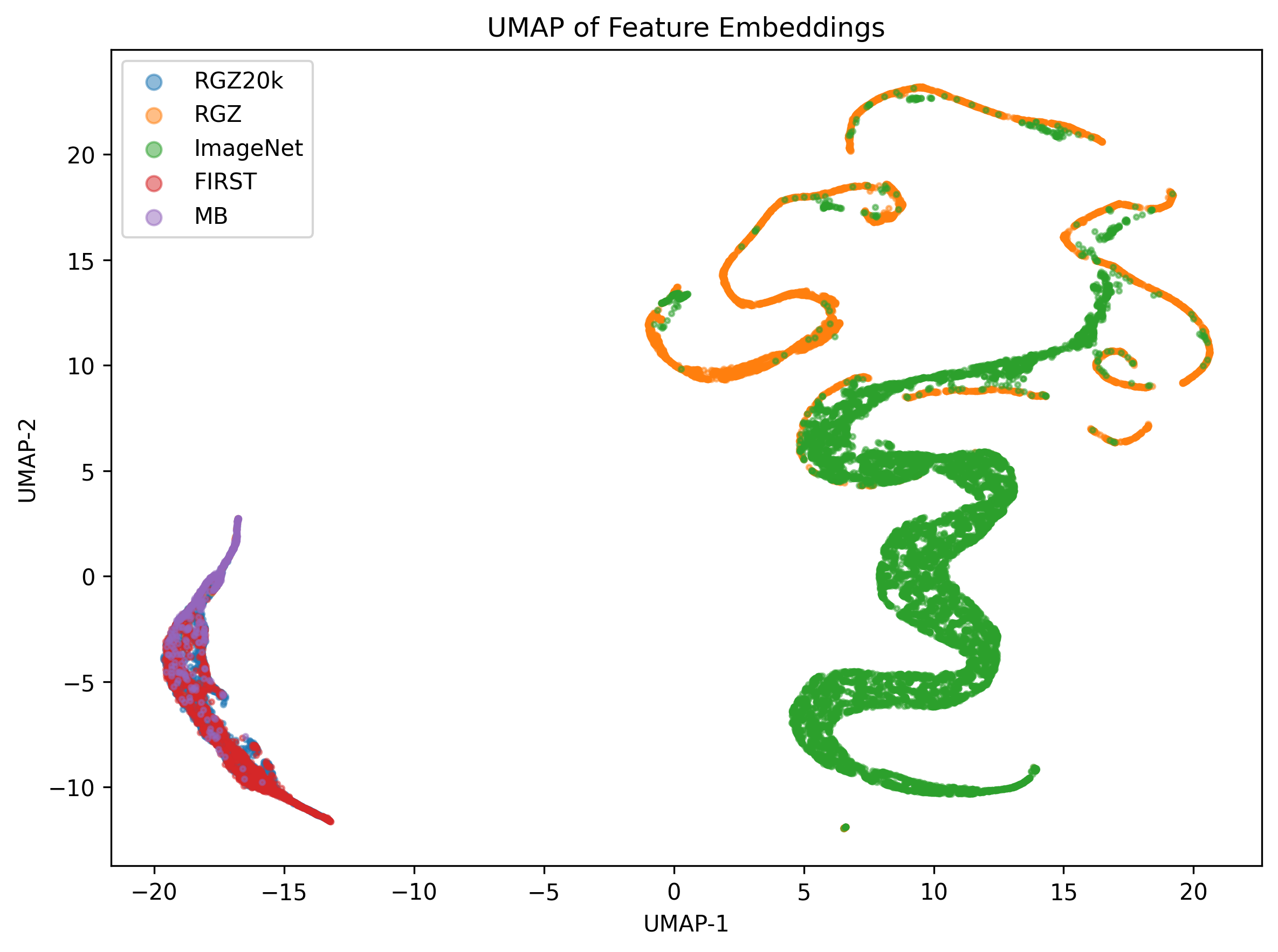}
    \includegraphics[width=0.48\linewidth]{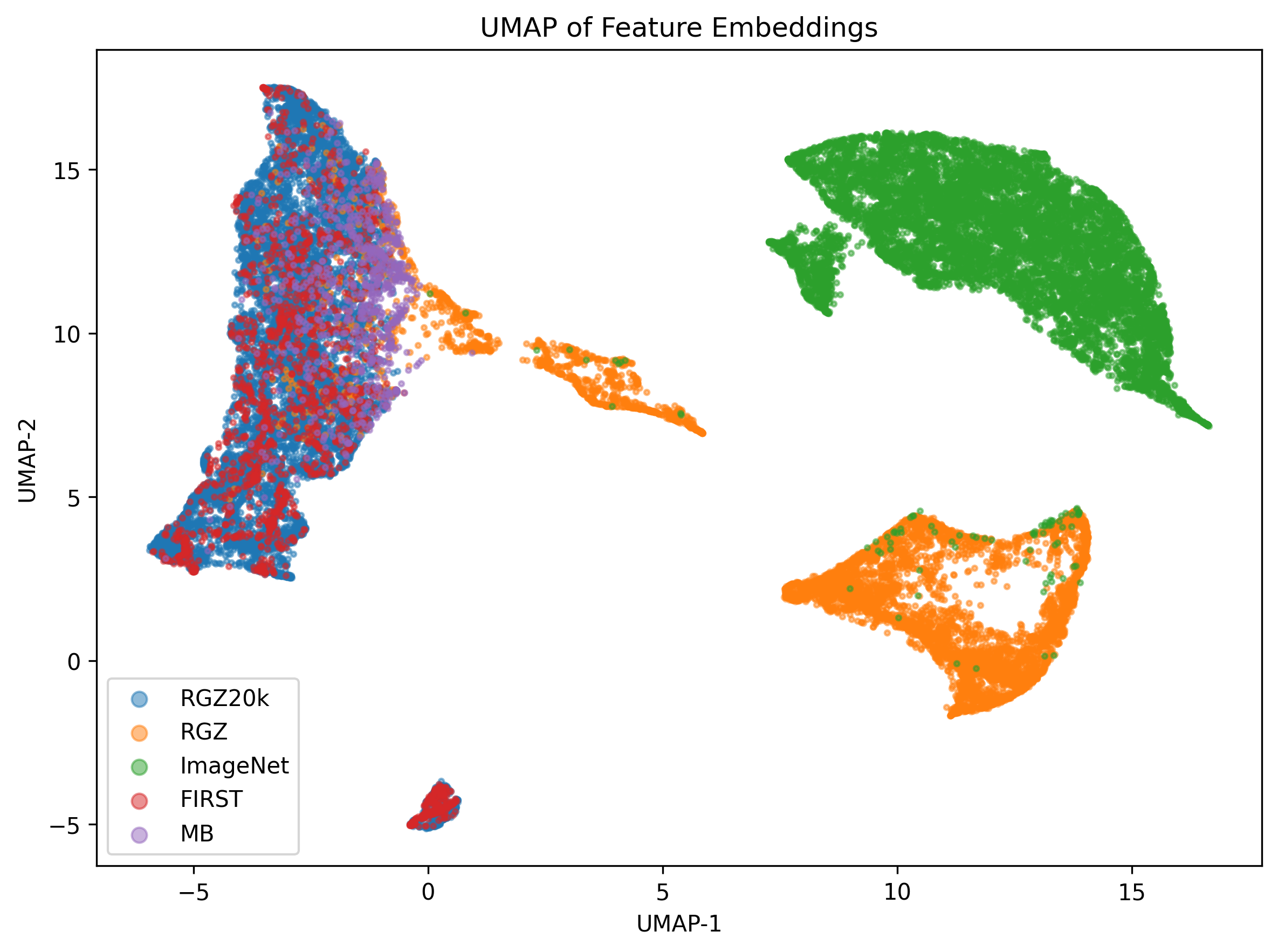}
    \caption{Two-dimensional UMAP projections of feature embeddings for our datasets, using BYOL (left) and SimCLR (right) pre-trained on RGZ20k.}
    \label{fig:byol}
\end{figure}

\begin{figure}[h]
    \centering
    \includegraphics[width=0.48\linewidth]{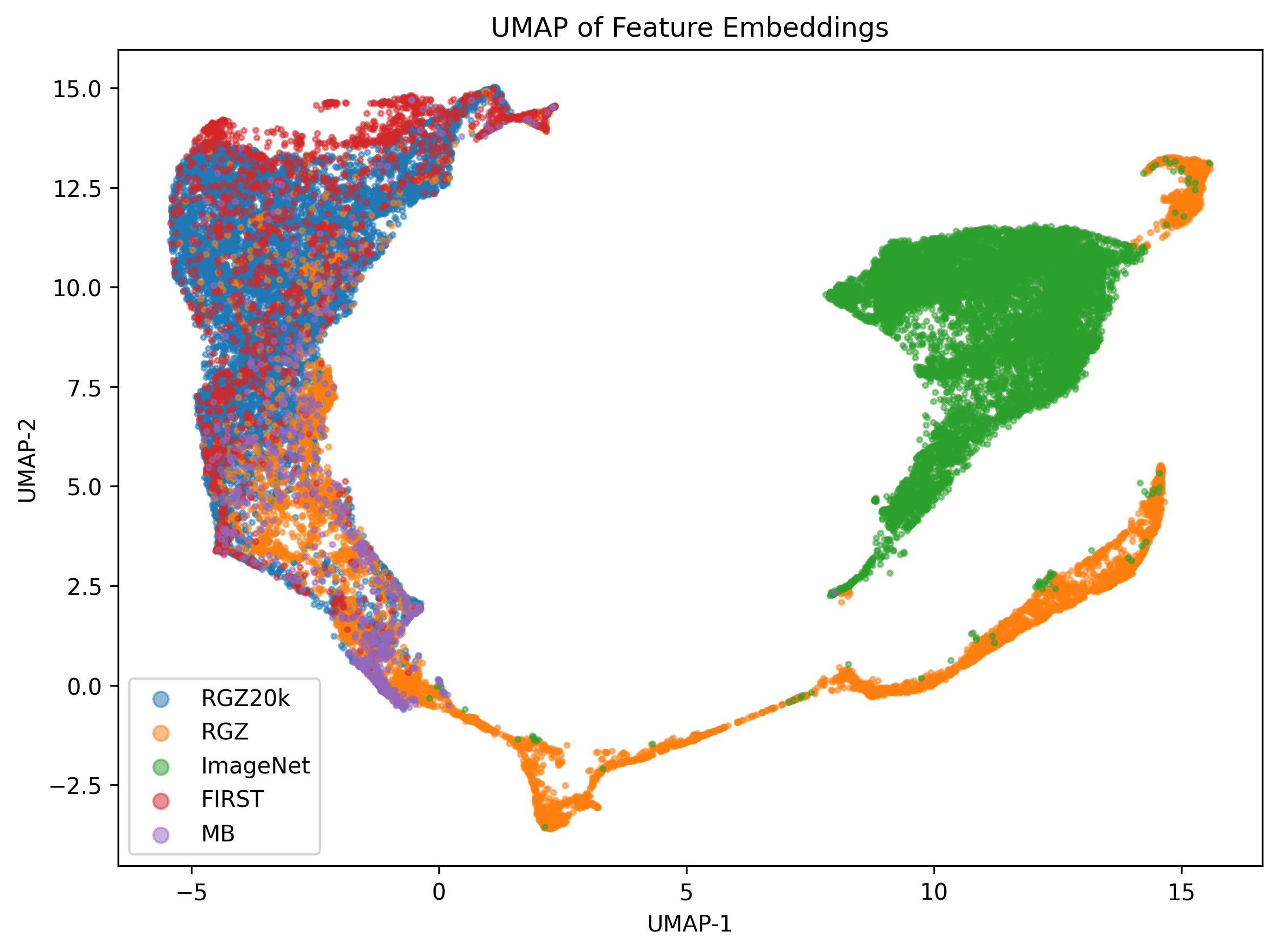}
    \includegraphics[width=0.48\linewidth]{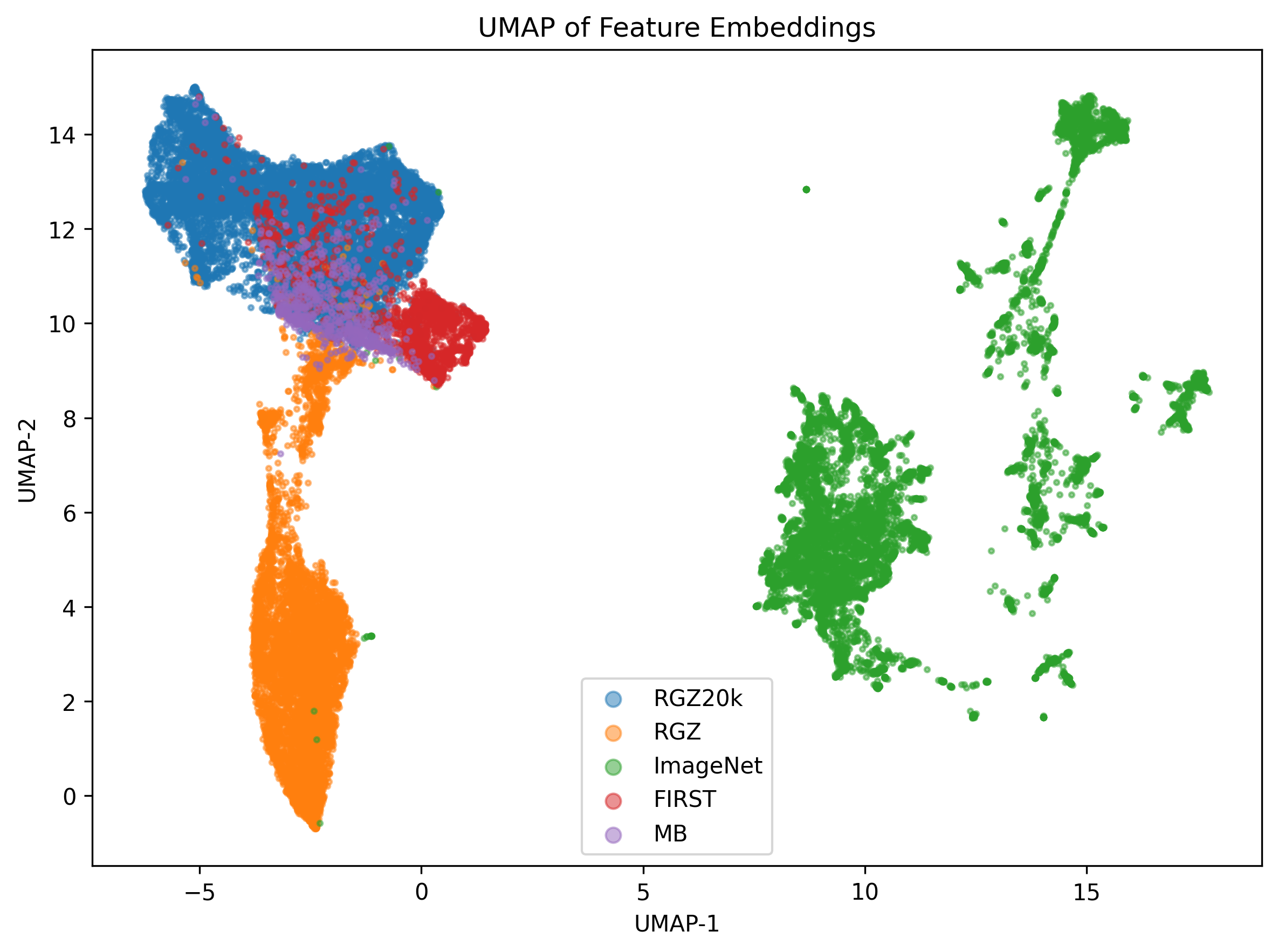}
    \caption{Two-dimensional UMAP projections of feature embeddings for our datasets, using EfficientNet pre-trained on RGZ20k via LeJEPA (left) and EUPE-ViT-Small (right).}
    \label{fig:enet}
\end{figure}

\begin{figure}[h!]
    \centering
    \includegraphics[width=0.48\linewidth]{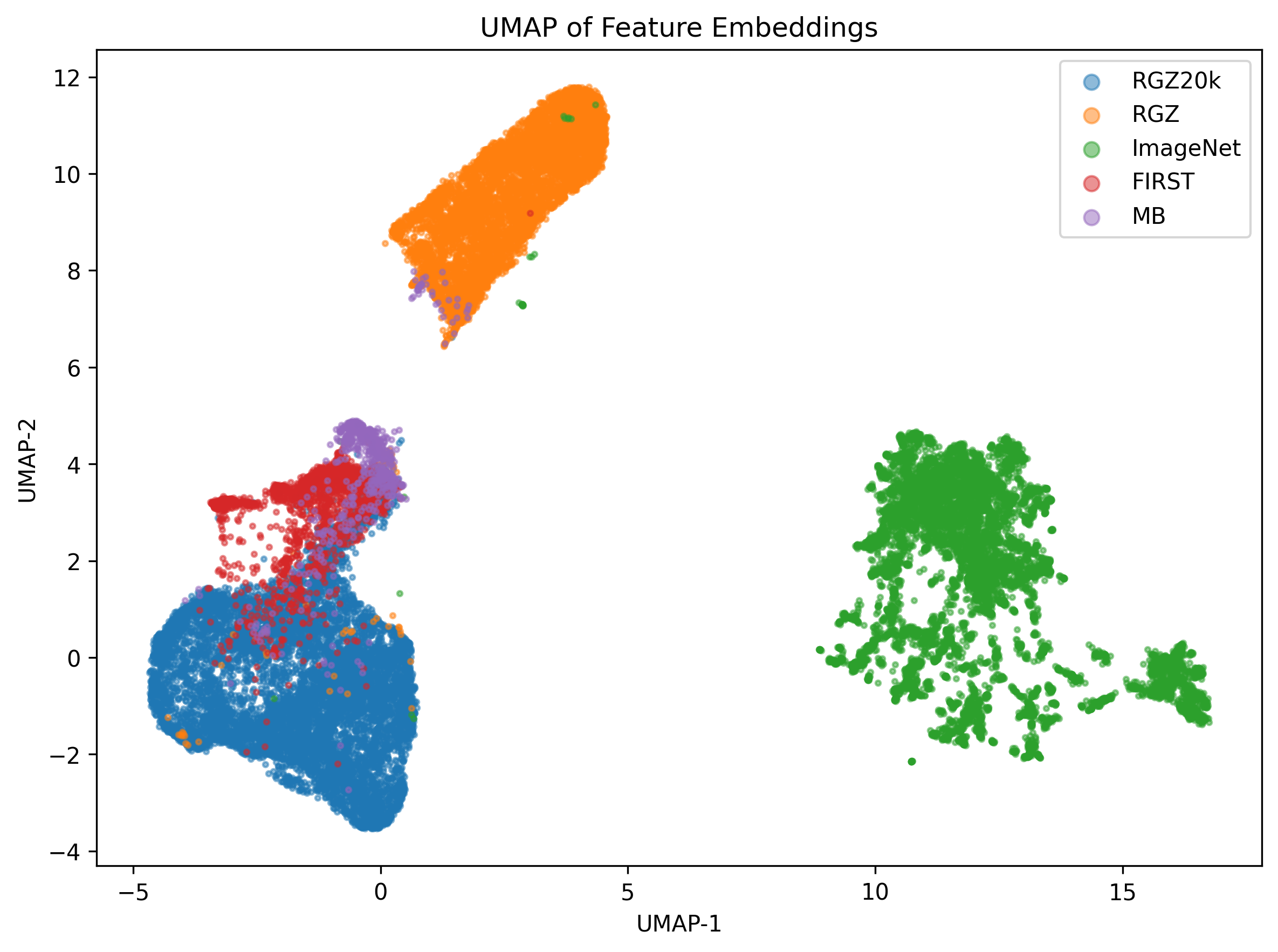}
    \caption{Two-dimensional UMAP projections of feature embeddings for our datasets, using DINOv3-Large.}
    \label{fig:dinov3}
\end{figure}

Figures \ref{fig:byol}, \ref{fig:enet} and \ref{fig:dinov3} show UMAPs of feature embeddings for all models (except those shown in Figure \ref{fig:umap}) and datasets used in this paper. DINOv3-Large is also shown for reference as a 300M-parameter vision foundation model. DINOv3 has a mean inference latency per image of 16.79 ms, making it almost 2x as slow to perform inference with this model as with ViT-Tiny.

\subsection{Scaling behavior}

\begin{figure}[h]
    \centering
    \includegraphics[width=\linewidth]{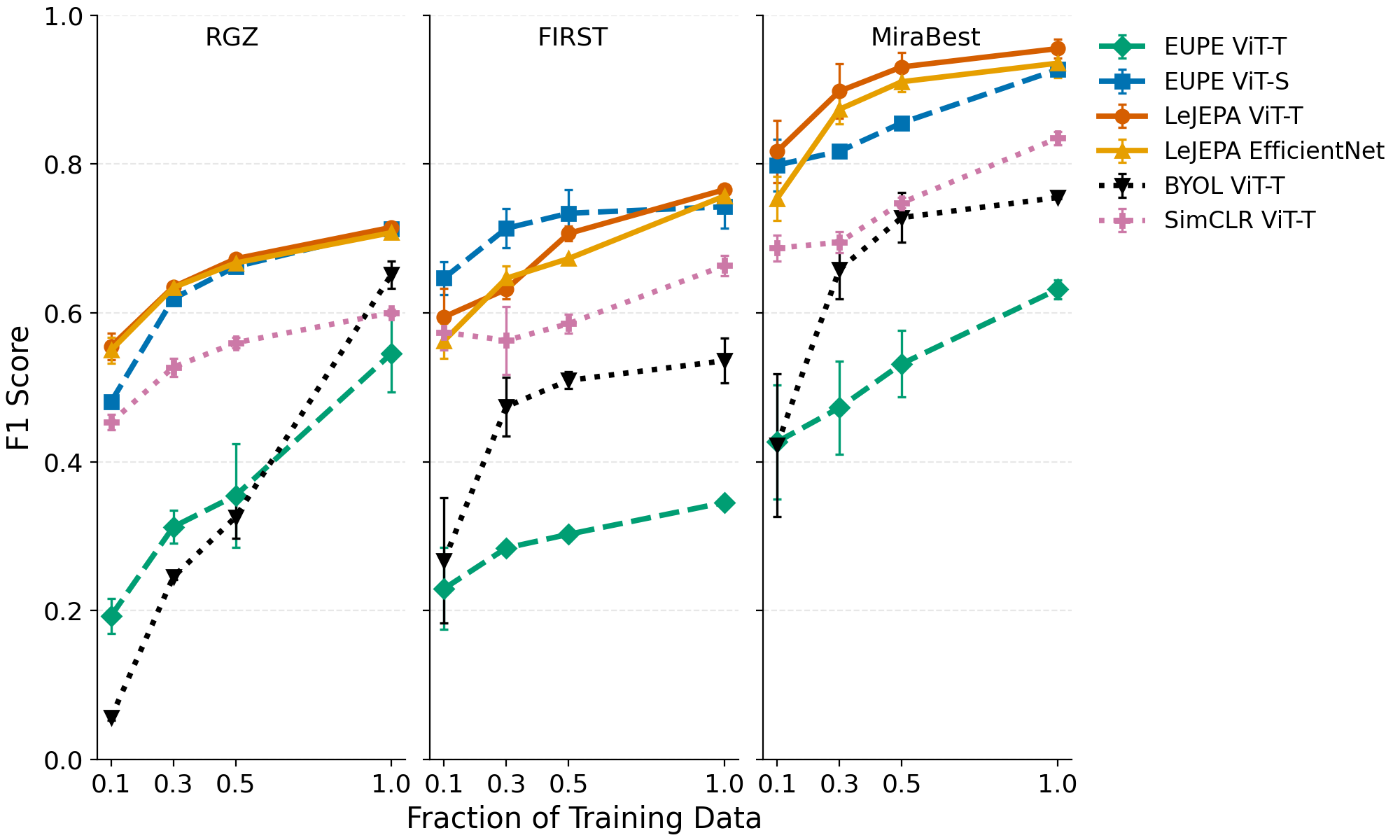}
    \caption{F1 score as a function of the fraction of available training data ($0.1, 0.3, 0.5,$ and $1.0$) on three radio galaxy datasets: RGZ (left), FIRST (middle), and MiraBest (right) . Error bars denote performance variation across runs with three different random seeds.}
    \label{fig:scaling}
\end{figure}

Figure \ref{fig:scaling} shows fully fine-tuned classification performance as it scales with the number of training samples. ViT-Tiny pre-trained via LeJEPA performed competitively with the EUPE ViT-Small model across all label regimes, despite operating with significantly fewer parameters. On RGZ and MiraBest, both LeJEPA models achieve comparable or higher F1 scores than EUPE ViT-S at every data fraction, reaching peak F1 scores near $0.95$ on MiraBest at full data. 

Compared to standard ViT-Tiny self-supervised baselines, LeJEPA maintains higher F1 scores across most label fractions. While BYOL ViT-T exhibits steep gains as the data fraction increases, its performance degrades severely in low-label regimes, dropping below $F1 = 0.10$ on RGZ and $F1 = 0.30$ on FIRST at $10\%$ data. SimCLR ViT-T demonstrates higher stability than BYOL under label scarcity (e.g., $F1 \approx 0.45$ on RGZ at $10\%$ data) but saturates at lower peak F1 scores relative to both LeJEPA variants and EUPE ViT-S across all three benchmarks.

\subsection{UMAPs of Evaluation Dataset Embeddings}\label{sec:umap_eval}

\begin{figure*}[p]
    \centering

    \begin{subfigure}{0.32\textwidth}
        \centering
        \includegraphics[width=\linewidth]{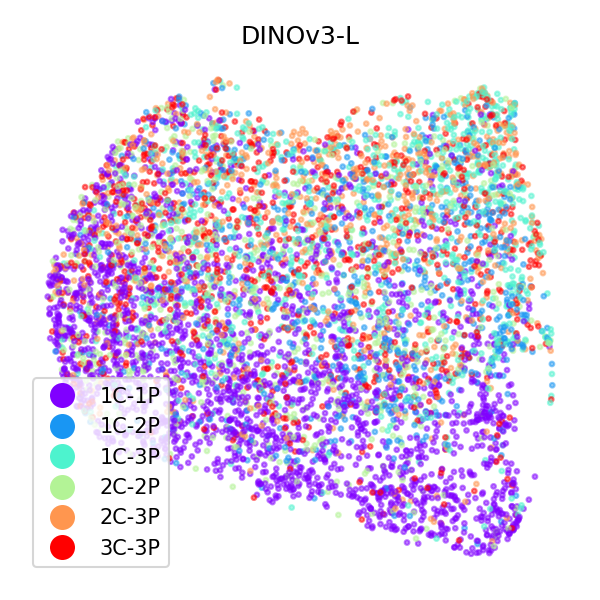}
    \end{subfigure}
    \hfill
    \begin{subfigure}{0.32\textwidth}
        \centering
        \includegraphics[width=\linewidth]{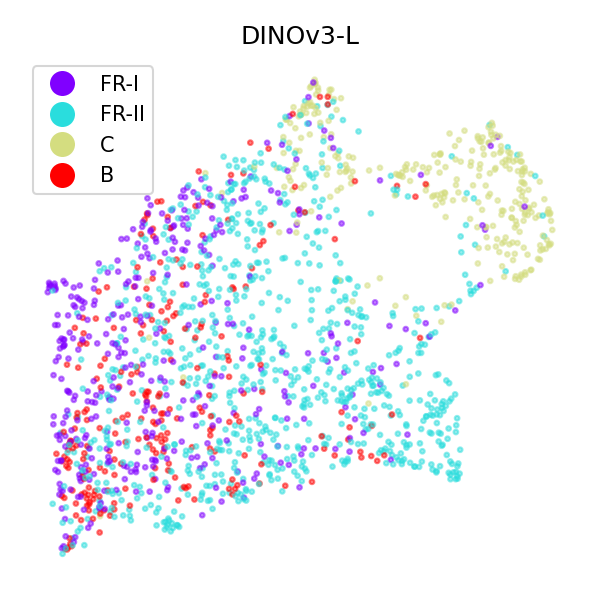}
    \end{subfigure}
    \hfill
    \begin{subfigure}{0.32\textwidth}
        \centering
        \includegraphics[width=\linewidth]{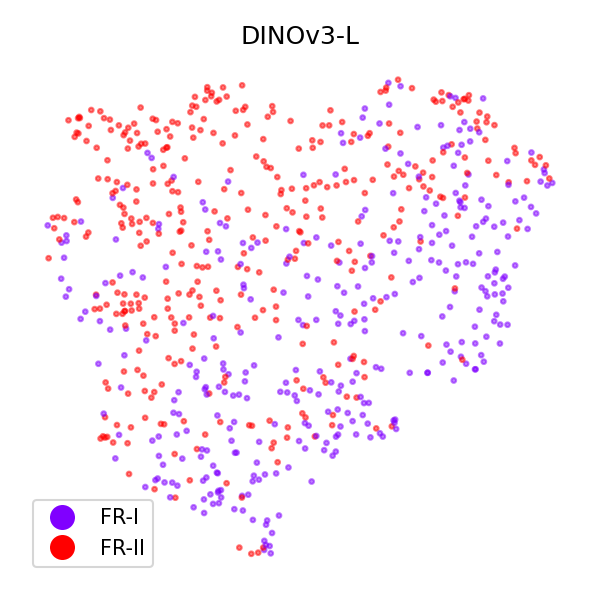}
    \end{subfigure}

    \vspace{0.5em}

    \begin{subfigure}{0.32\textwidth}
        \centering
        \includegraphics[width=\linewidth]{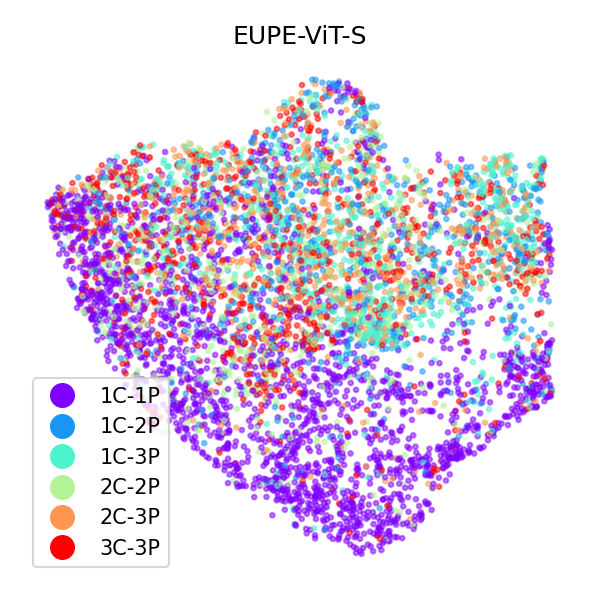}
    \end{subfigure}
    \hfill
    \begin{subfigure}{0.32\textwidth}
        \centering
        \includegraphics[width=\linewidth]{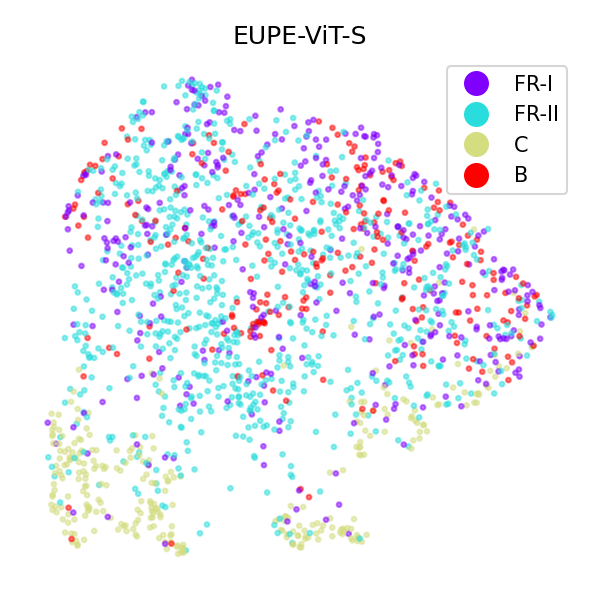}
    \end{subfigure}
    \hfill
    \begin{subfigure}{0.32\textwidth}
        \centering
        \includegraphics[width=\linewidth]{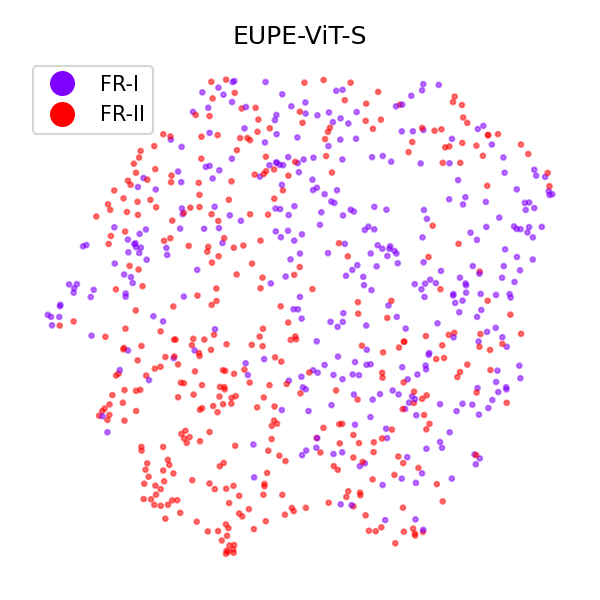}
    \end{subfigure}

    \begin{subfigure}{0.32\textwidth}
        \centering
        \includegraphics[width=\linewidth]{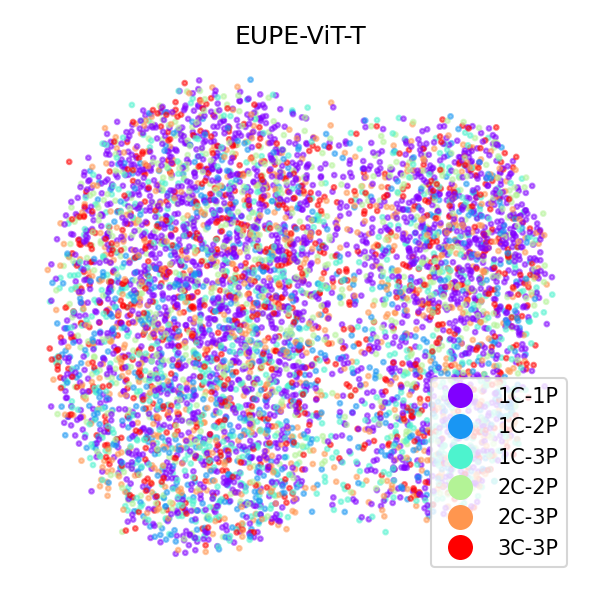}
    \end{subfigure}
    \hfill
    \begin{subfigure}{0.32\textwidth}
        \centering
        \includegraphics[width=\linewidth]{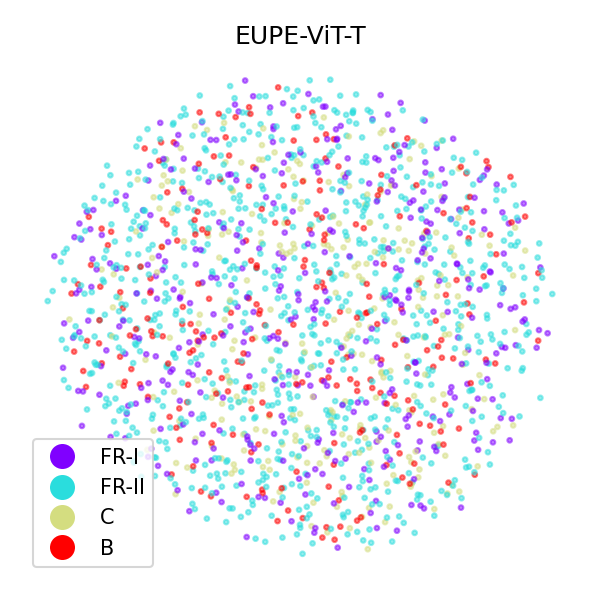}
    \end{subfigure}
    \hfill
    \begin{subfigure}{0.32\textwidth}
        \centering
        \includegraphics[width=\linewidth]{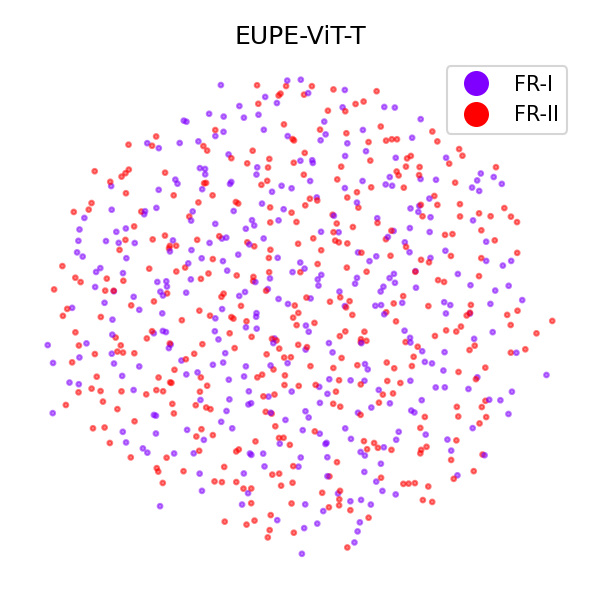}
    \end{subfigure}

    \begin{subfigure}{0.32\textwidth}
        \centering
        \includegraphics[width=\linewidth]{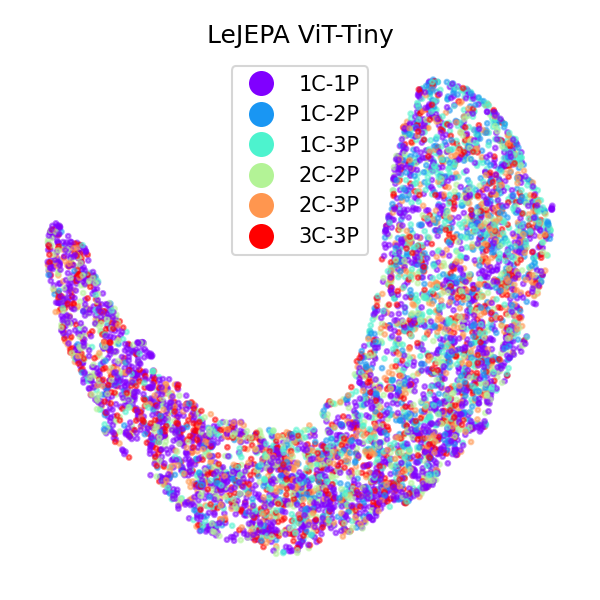}
    \end{subfigure}
    \hfill
    \begin{subfigure}{0.32\textwidth}
        \centering
        \includegraphics[width=\linewidth]{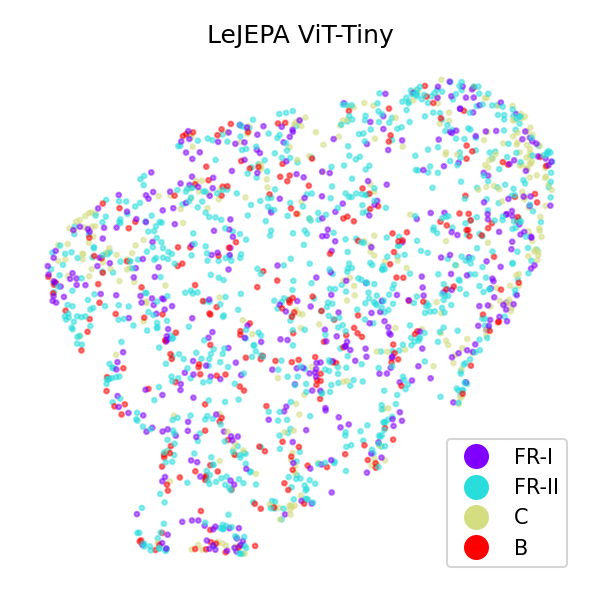}
    \end{subfigure}
    \hfill
    \begin{subfigure}{0.32\textwidth}
        \centering
        \includegraphics[width=\linewidth]{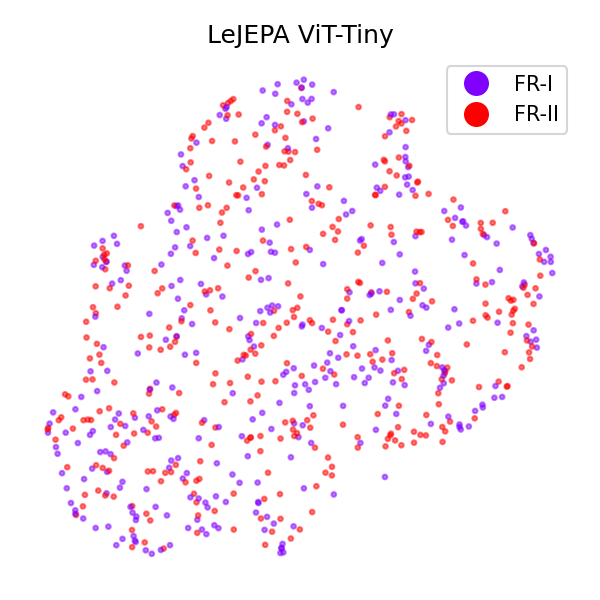}
    \end{subfigure}

    \caption{UMAPs of feature embeddings for models and evaluation datasets used in this study. The left-most column shows the 6-class RGZ dataset, the center shows the 4-class FIRST dataset and the right column shows the binary MiraBest dataset.}
    \label{fig:full_umap_grid}
\end{figure*}

\begin{figure*}[p]
    \centering

    \begin{subfigure}{0.32\textwidth}
        \centering
        \includegraphics[width=\linewidth]{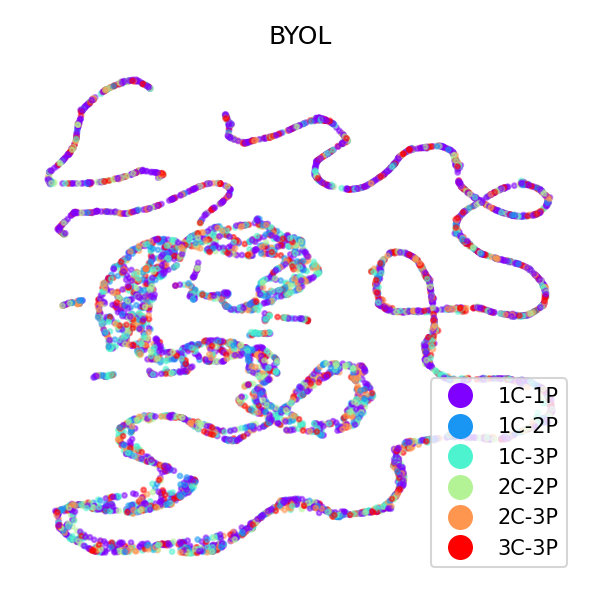}
    \end{subfigure}
    \hfill
    \begin{subfigure}{0.32\textwidth}
        \centering
        \includegraphics[width=\linewidth]{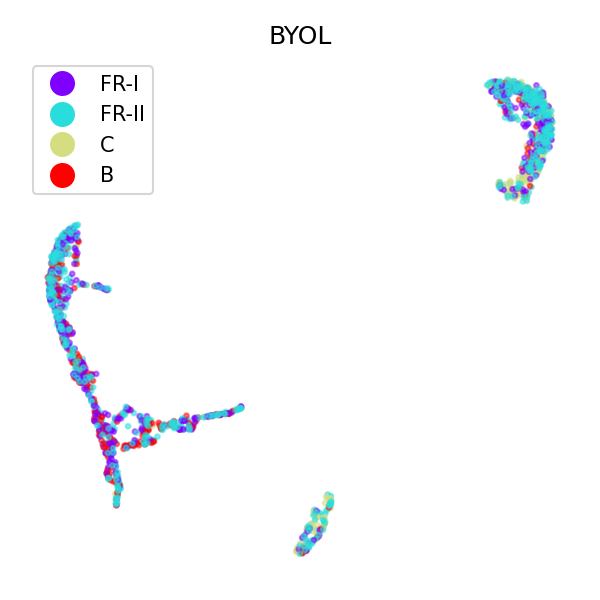}
    \end{subfigure}
    \hfill
    \begin{subfigure}{0.32\textwidth}
        \centering
        \includegraphics[width=\linewidth]{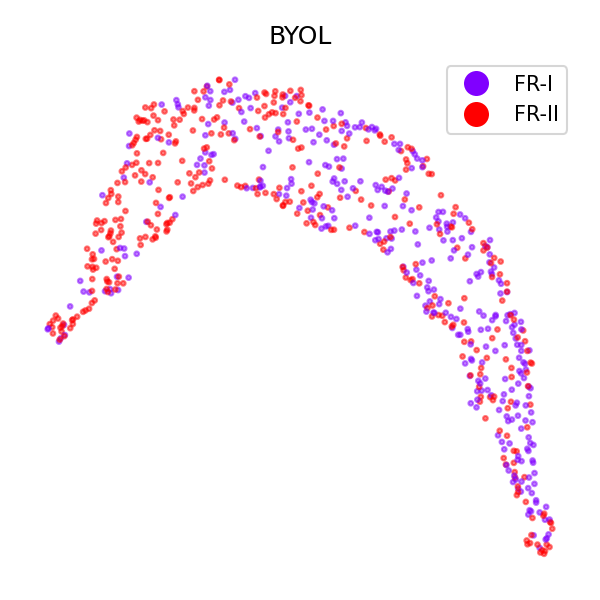}
    \end{subfigure}

    \vspace{0.5em}

    \begin{subfigure}{0.32\textwidth}
        \centering
        \includegraphics[width=\linewidth]{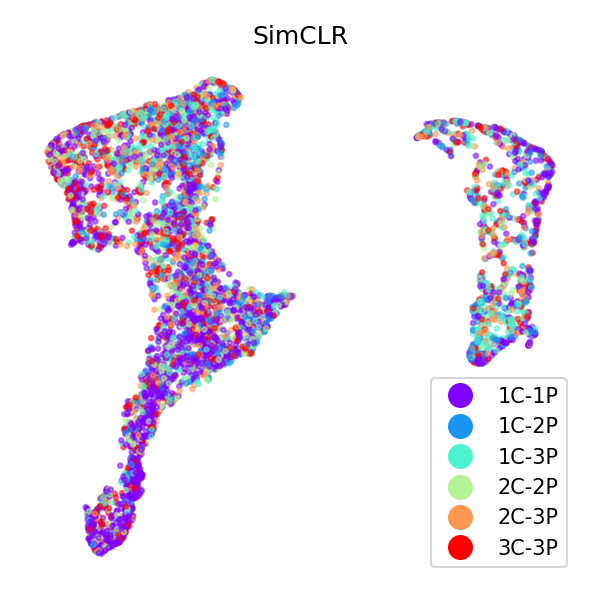}
    \end{subfigure}
    \hfill
    \begin{subfigure}{0.32\textwidth}
        \centering
        \includegraphics[width=\linewidth]{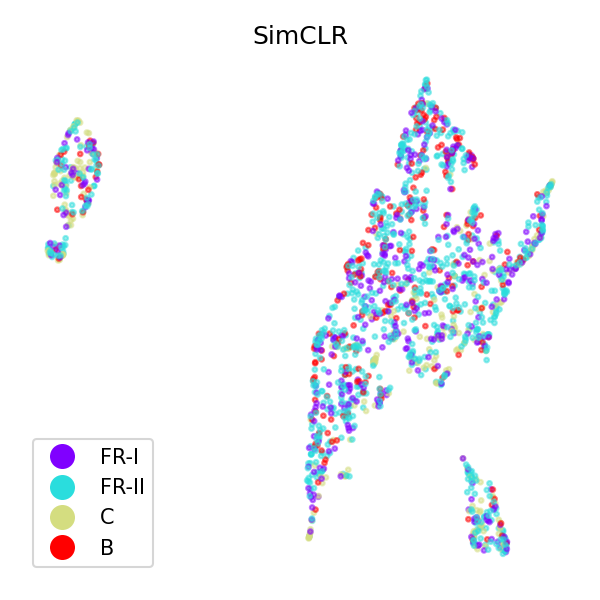}
    \end{subfigure}
    \hfill
    \begin{subfigure}{0.32\textwidth}
        \centering
        \includegraphics[width=\linewidth]{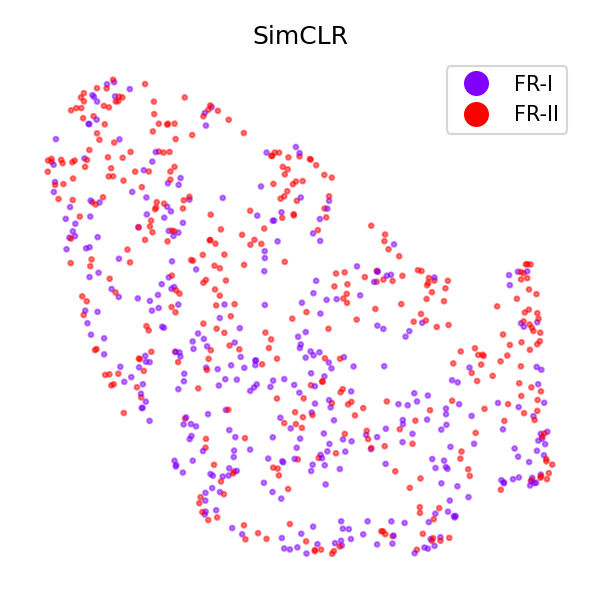}
    \end{subfigure}

    \begin{subfigure}{0.32\textwidth}
        \centering
        \includegraphics[width=\linewidth]{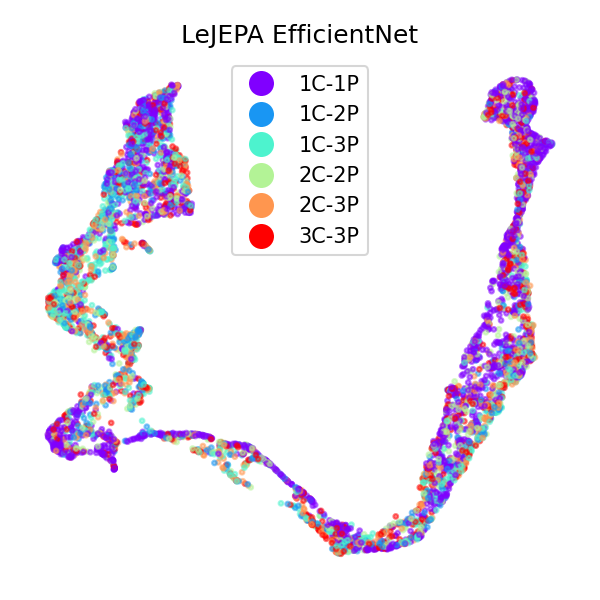}
    \end{subfigure}
    \hfill
    \begin{subfigure}{0.32\textwidth}
        \centering
        \includegraphics[width=\linewidth]{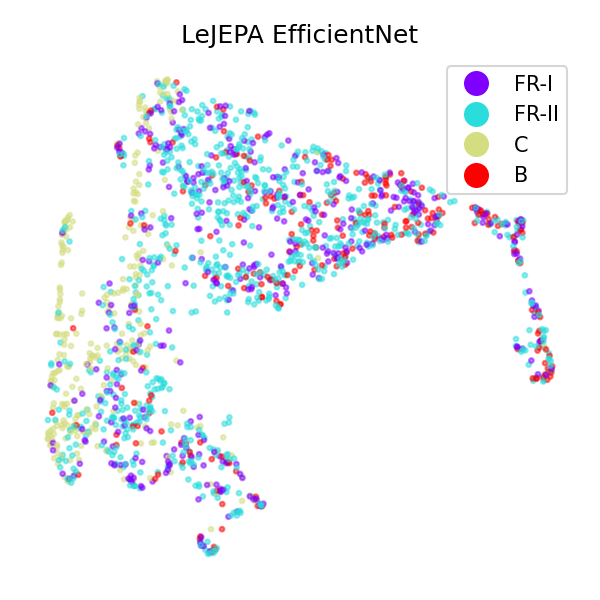}
    \end{subfigure}
    \hfill
    \begin{subfigure}{0.32\textwidth}
        \centering
        \includegraphics[width=\linewidth]{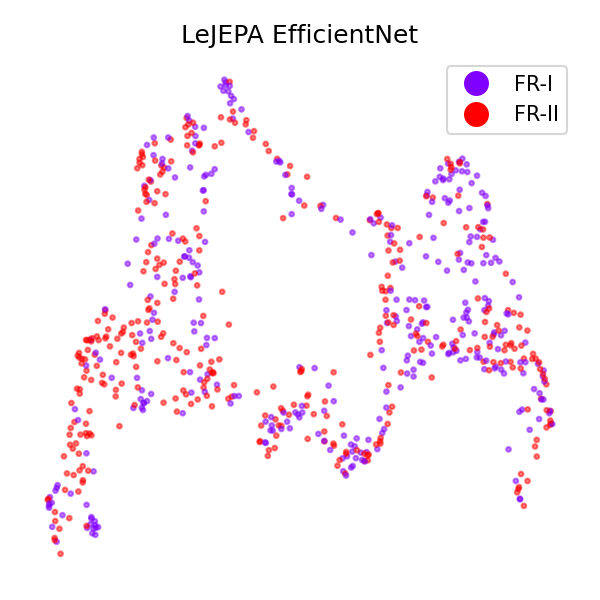}
    \end{subfigure}

    \caption{UMAPs of feature embeddings for models and evaluation datasets used in this study. The left-most column shows the 6-class RGZ dataset, the center shows the 4-class FIRST dataset and the right column shows the binary MiraBest dataset.}
    \label{fig:full_umap_grid2}
\end{figure*}

Figures \ref{fig:full_umap_grid} and \ref{fig:full_umap_grid2} show UMAPs displaying the embedding space of each evaluation dataset for each model used in this study, with the addition of DINOv3-Large for more context. Classes are labeled as indicated in Table \ref{tab:datasets}. RGZ classes are named for the number of radio components (C) and the number of observed flux brightness peaks (P).

As these embeddings are extracted from models that have not undergone task-specific fine-tuning, we do not expect obvious visual separation between classes. Classification results clearly show that following fine-tuning or even a linear probe, representations are in fact class-separable (with the exception of EUPE-ViT-T). 

\end{document}